# Title: The microscopic origin of abrupt transitions in interdependent systems

**Authors:** Bnaya Gross[1,2*†], Irina Volotsenko[3†], Yuval Sallem[3],

Nahala Yadid[3], Ivan Bonamassa[4], Shlomo Havlin[3], Aviad Frydman[3**]

**Affiliations:** [1]Network Science Institute, Northeastern University, Boston, MA 02115

[2]Department of Physics, Northeastern University, Boston, MA 02115

[3]Department of Physics, Jack and Pearl Resnick Institute, Bar-Ilan University, 52900 Ramat-Gan, Israel

[4]Department of Network and Data Science, CEU, Quellenstrasse 51, A-1100 Vienna, Austria

[*]Corresponding author. Email: bnaya.gross@gmail.com

[**]Corresponding author. Email: aviad.frydman@gmail.com

[†]These authors contributed equally to this work.

**Abstract: Phase transitions are fundamental features of statistical physics. While the well-studied continuous phase transitions are known to be controlled by external *macroscopic* changes in the order parameter, the origin of abrupt transitions is not yet clear. Here we show that abrupt phase transitions may occur due to unique internal *microscopic* cascading mechanism, resulting from dependency interactions. We experimentally unveil the underlying mechanism of the abrupt transition in interdependent superconducting networks to be governed by a unique metastable state of a long-living resistance cascading plateau. This plateau is characterized by spontaneous *microscopic* changes that last for *thousands* of seconds, followed by a *macroscopic* phase shift of the system. Similar microscopic mechanisms are expected to be found a variety of systems showing**

**abrupt transitions.**

**Main Text:** Phase transitions (PTs) are among the most intriguing phenomena of statistical mechanics and are usually classified by their macroscopic features close to the critical point (*1–3*). Second-order PTs are characterized by a continuous transition of the order parameter at the critical point, while first-order PTs display an abrupt change between phases (*4*). In contrast to first-order PTs, second-order PTs display critical behavior near the critical point characterized by the scaling laws of different quantities (*2, 3*). A unique class of PTs is mixed-order PTs, which display both an abrupt change similar to first-order transitions and scaling laws and critical exponents near the critical point similar to second-order transitions (*5, 6*). Here we present experimental findings on the microscopic mechanism origin of interdependent superconducting networks that exhibit a mixed-order phase transition.

The paradigm of physical interdependent networks has recently been introduced by the realization of the first physical manifestation of an experimental setup of interdependent systems - interdependent superconducting networks (ISN) (*7*). This experimental discovery paves the way for a new research avenue of interdependent materials. ISNs have been found to experience an abrupt transition, as predicted by the theory of interdependent networks (*8*) resulting from the two types of interactions in the system: connectivity interactions within each network allowing current to flow, and dependency interactions between the networks in the form of thermal dissipation. Nonetheless, the underlying mechanism of the abrupt transition and its critical behavior remained elusive. Here we reveal experimentally and theoretically the underlying mechanisms behind the abrupt jump observed in ISNs. We identify and characterize the temporal-spatial scaling of the long-term plateau state which represents a novel *microscopic* mechanism of PTs. Like in epidemic spreading, these microscopic changes are characterized by a branching factor which equals exactly *one* at the critical point and below (above) *one* above (below) criticality. We hypothesize that a novel universality class defines the critical behavior of physical interdependent networks, as found here in our characterization of the critical behavior of ISNs.

# Results

## Experimental Setup

Our ISN system is composed of two coupled amorphous indium oxide (a − InO) disordered $2D$ superconducting networks, each containing a grid of $\mathcal{N} = L \times L$ segments with $L$ being the linear size of the system (see Fig. S1). The two layers are separated by an electrically insulating spacer ($Al_2O_3$) that has finite heat conductance. The interdependent networks paradigm is manifested here by the *connectivity* links, represented by currents within each network layer, and *dependency* links realized via heat transfer between the layers. Each connectivity link $i$ is characterized by a Josephson junction $I - V$ characteristics (*9*) with a distinct critical temperature $T_{c,i}$ and a critical current $I_{c,i}$ determining its superconductor-normal (S-N) transition threshold (*10–20*).

## Resistance versus Current Measurements

The system is set at a fixed and controllable temperature and the network resistance, $R_N$, is experimentally measured as we sweep the bias current $I_b$ from high to low and back. During this process, an abrupt jump is observed at a critical point $I_{c,\leftarrow}$ from the mutual normal state (N) to the mutual superconducting state (S) as the current decreases, and from the S-state back to the N-state at $I_{c,\rightarrow}$ as the current increases, showing a hysteresis phenomenon, see Fig. 1**a**. This result can be obtained theoretically by numerically solving the Kirchhoff equations of both networks simultaneously while accounting for the thermal coupling between the layers (*7*) as seen in Fig. 1**b**. The characteristics of ISNs clearly exhibit an abrupt transition but also a critical exponent $\beta$, describing the scaling of $R_N$ as the current approaches the critical point $I_c$ from the N-state to the S-state as

$$\Delta R \sim (\Delta I)^\beta. \tag{1}$$

where $\Delta I = I_b - I_c$. Both, the experimental and numerical simulation results follow the scaling behavior with the same value of the critical exponent $\beta = 1/2$ (Figs. 1**a,b** insets respectively). The same exponent is observed for fixed current and varied temperature (Fig. S2) and in simulations of other interdependent systems including abstract percolation on interdependent networks (*21*) and interdependent ferromagnetic networks (*22*), suggesting that the transition is universal in mixed-order PTs.

The role of the dependency between the superconducting networks formed by the heat dissipation in ISNs is intrinsic to the critical behavior and can be understood by the I − V curves for different system sizes as shown in Figs. 2**a,d**. Small systems produce weak heat dissipation, making the coupling weak, thus resulting in a continuous transition similar to the transition observed in single isolated networks (*7*). Large systems, on the other hand, dissipate enough heat for an abrupt transition to emerge, suggesting that the sample size also influences the critical behavior. Furthermore, as seen in Figs. 2**b,e**, the critical transition points are found to depend on the system size. While for small systems the transition is continuous and $I_{c,\rightarrow} = I_{c,\leftarrow}$, for large systems hysteresis is observed showing $I_{c,\rightarrow} > I_{c,\leftarrow}$ and both critical thresholds increase with the system size as shown in Figs. 2**b,e**. The hysteresis presents three regimes: N-state for $I_b > I_{c,\rightarrow}$, S-state for $I_b < I_{c,\leftarrow}$ and an S/N-state depending on the initial conditions for $I_{c,\leftarrow} < I_b < I_{c,\rightarrow}$. The increase of the critical current with the system size allows us to estimate the correlation length critical exponent $\nu$ using the scaling relation (*23–25*)

$$I_c(\infty) - I_c(L) \sim L^{-1/\nu} \tag{2}$$

where $\nu$ is the correlation length exponent and $I_c(\infty)$ is a fitting parameter (see supplementary text). Figs. 2**c,f** show for both experimental and theoretical results the critical exponent $\nu = 3/4$, as predicted for interdependent systems (*26*).

## Dynamics During the Transition

Identifying the underlying physical mechanisms of a phase transition is essential to understand its nature. To this end, we measure experimentally the dynamics of the system, i.e. the resistance versus time of the system, $R(t)$, *during* the abrupt transition from the mutual N-state to the mutual S-state at and near the relevant critical threshold $I_c$, see Fig. 1. The results shown in Fig. 3**a** are based on the following protocol: (i) The system is controlled and fixed at a temperature $T < T_c$ and the bias current is set at $I_b > I_c$ so the system is tuned to be in the N-state. (ii) At time $t = 0$ the current is abruptly switched to a value $I_b < I_c$ and $R(t)$ is measured. Remarkably, when the bias current is switched very close to the critical current $I_b \rightarrow I_c$, the resistance curves display a semi-saturated resistance for a long time (long-term plateau) before it drops abruptly to zero resistance as the system transits to the S-state (Fig. 3**a,c** show experimental and numerical results respectively). The time, $\tau$, of this *plateau* can last for *thousands* of seconds. That is macroscopic times many orders of magnitudes longer than the timescales of electronic interactions ($\tau_e \sim 10^{-12} - 10^{-10} s$) within each network or phonon between them ($\tau_p \sim 10^{-8} - 10^{-6} s$), indicating that the interplay between the layers plays a significant role in the process. The inset of Fig. 3**a** zoom in on the plateau regime, showing that the resistance does not monotonically decrease with time but fluctuates. Similar behavior are reproduced by the theory upon introducing thermal fluctuations (Fig. 3**c**).

We repeat the above process by deep quenching to slightly lower values of $I_b$ and find that, as $I_b$ decreases, $\tau$ decreases as well (Fig. 3**a**). The measured plateau duration time $\tau$ is found to follow a scaling behavior with the distance of the bias current from the critical current as

$$\tau \sim (\Delta I)^{-\zeta} \tag{3}$$

with an exponent consistent with $\zeta = 1/2$ as seen in Figs. 3**b,e** for experimental and numerical results respectively. This value is consistent with the theoretical prediction for percolation on abstract interdependent networks (*27*).

## Microscopic Mechanism Origin

The extremely long plateau indicates a mechanism of spontaneous long-term microscopic changes. This long-term plateau suggests that, due to the long-range dependency in such ISNs, each segment in one network may influence every segment in the second network. Thus, the interactions become nearly spatially random where, near criticality, phase changes (N-S) of elements are generated anywhere in the system depending on $T_{ci}$ and $I_{ci}$ of the individual segments. In this near criticality regime, the phase transition is governed by a spontaneous microscopic *random-cascading*, i.e. a chain of events where one element, that has changed its phase, influences on average one element at a different location with closest $T_{ci}$ and $I_{ci}$ in the other network due to thermal dissipation. This process continues until enough elements undergo a phase-change thus causing a macroscopic PT. We note that, while the experiment focuses on the transition from N to S, an analogous process is expected to occur for a transition from S to N, as illustrated in Fig. 4.

Since the changes during the plateau are microscopic, in order to influence the whole system, we expect $\tau$ to depend on the sample size. For this, we measure the time duration of the plateau, $\tau$, for different system sizes. Indeed, Fig. 3**b,d** shows experimental and theoretical (based on numerical solutions) measurements of the dependence of $\tau$ on the system size which follows

$$\tau \sim \mathcal{N}^{\psi} \tag{4}$$

with $\psi = 1/3$ which satisfies the relation $\psi = \zeta/(\nu d)$ (see Supplementary Text). Note that this exponent has also been observed in percolation on abstract interdependent networks (*27*). This result supports our paradigm that the plateau, during the abrupt transition, represents a *microscopic* process that occurs for a *macroscopic* time. The collapse of all experimental curves in the plot of Fig. 3**b** is rather striking and further supports the interpretation of the slowing down resulting from critical branching.

To further support this microscopic paradigm we consider the branching factor, $\mathcal{R}$, which defines the average number of segments in a network that are affected by a single segment in the other network that changes their phase e.g, from N to S. This parameter is analogous to the $\mathcal{R}$ in a pandemic when grading the Covid-19 or any disease's ability to spread. It measures how many people, on average, an infected individual will infect before recovery or death. An $\mathcal{R} < 1$ means the disease is being suppressed while an $\mathcal{R} > 1$ means it is increasing exponentially fast. The case of $\mathcal{R} = 1$ is critical and means that a person with the disease passes it onto only one other person on average and the population of infected increases linearly over time. Nevertheless, the systems are very different. In a pandemic, the PT is continuous, and $\mathcal{R} = 1$ is not stable over time. In our ISNs, on the other hand, the PT at criticality is abrupt and characterized by a long-term plateau, (which does not exist in epidemics) while during this abrupt transition, $\mathcal{R}$ is exactly *one* along the plateau (Fig. 3**a**), before the system abruptly changes its phase. This is because the heat dependency interaction range extends over the entire sample and hence, each segment that changes phase from N to S will affect on average exactly one segment in the other network with the closest $T_c$ and $I_c$. This implies that above $I_c$ it is expected that $\mathcal{R} < 1$ and below $I_c$ one can expect $\mathcal{R} > 1$ while the *critical branching* occurs at $\mathcal{R}_c = 1$.

In order to test the above hypothesis we quantify $\mathcal{R}$ in our ISNs along the plateau using the following procedure:

$$\mathcal{R}(t) = [R(t + \Delta_t) - R(t)]/[R(t) - R(t - \Delta_t)]. \tag{5}$$

for each network. Fig. 5 shows both experimentally and theoretically the average branching $\langle \mathcal{R} \rangle$

during the plateau for different currents around the critical point. Above the critical point ($\Delta I > 0$), the average branching factor $\langle \mathscr{R} \rangle$ is smaller than one leading to an early stop of the cascading process and the system remains at the N-state. Below the critical point ($\Delta I < 0$), the average branching factor is larger than one and the cascading process, transitions the system into the S-state. Exactly at the critical point ($\Delta I = 0$) a critical branching factor of $\langle \mathscr{R} \rangle_c = 1$ is observed. This behavior of the branching factor around criticality further supports the hypothesis of the spontaneous microscopic cascading during the abrupt mixed-order transition and we expect similar behavior in other interdependent systems. The experimental measurements of the plateau used to estimate the branching factor are shown in Fig. S3. Similar behavior for the branching factor is also observed for fixed current and varied temperature (Fig. S4).

## Conclusions

One of the most important aspects of phase transitions is its underlying mechanism which characterizes its nature. The state-of-the-art underlying mechanism of phase transitions usually involves *macroscopic* interventions such as changing an external parameter of the entire system for second-order transitions or the spreading of growing nucleating droplets in first-order transitions (*1–3*). Here we reveal that an abrupt *macroscopic* phase transition can occur due to spontaneous long-term *microscopic* changes whose time scale is macroscopically long and depends on the system size. This finding fundamentally alters our understanding of phase transitions and is expected to be found in a large class of systems. The features of the transition in our ISNs, characterized by the set of critical exponents $(\beta, \nu) = (1/2, 3/4)$ and $(\zeta, \psi) = (1/2, 1/3)$, is similar to that found for abstract percolation on interdependent networks (*21, 27*) and for the theoretical model of interdependent ferromagnetic networks (*6, 22, 28*) indicating that properties of mixed-order transitions may be universal and suggest that the microscopic mechanism could be the origin of mixed-order transitions observed in a wide range of different systems.

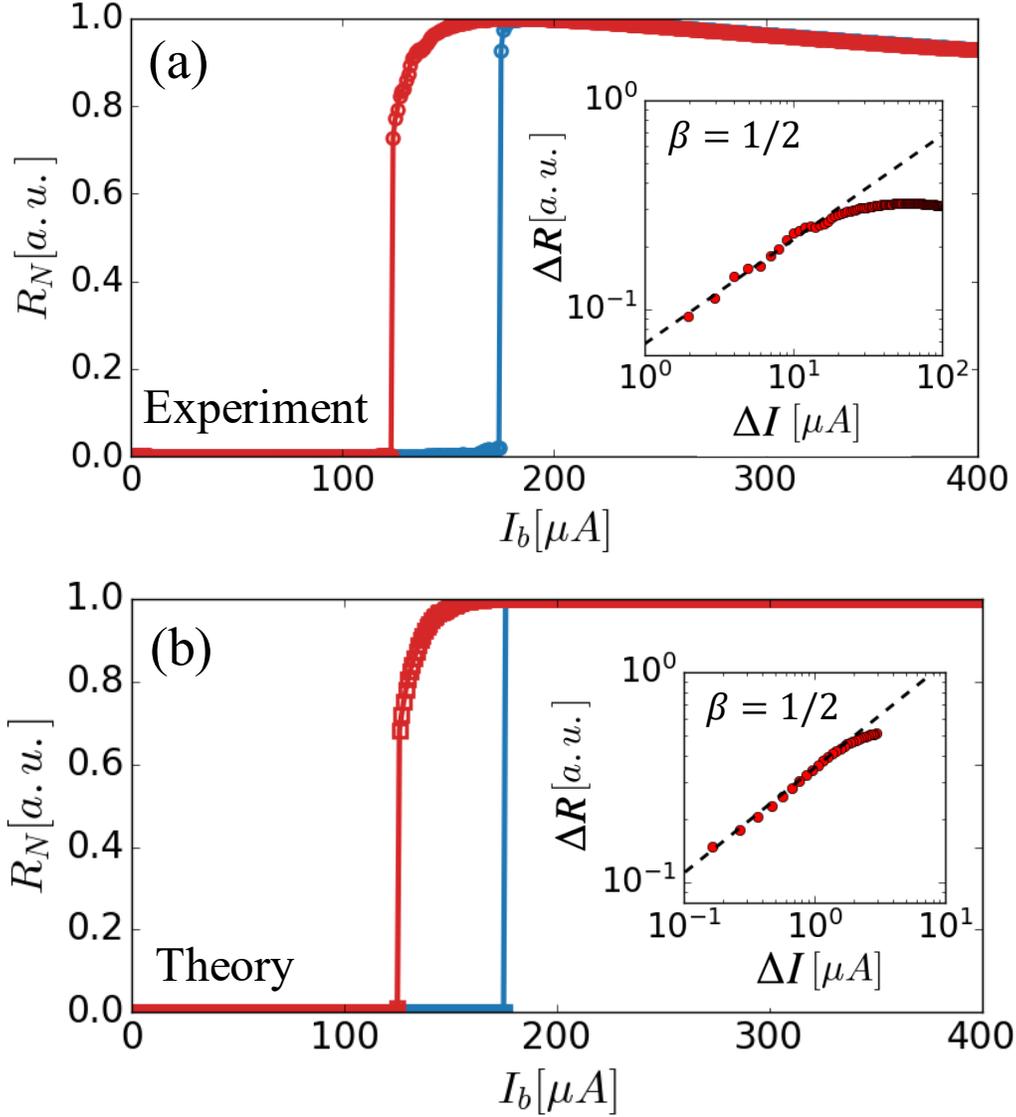

**Figure 1**: **Mixed-order transitions in interdependent superconducting networks.** For a fixed temperature, the network resistance, $R_N$, is measured for increasing current (blue) and decreasing current (red) showing hysteresis in both **(a)** experiment (for the top network of an $L = 416$ ISN) and **(b)** theory via numerical solution of the Kirchhoff equations. For decreasing current, the critical exponent $\beta = 1/2$ (Eq. (1)) is observed near the critical point $I_c$ from N to S in both experiment and numerical solution (insets **(a)** and **(b)** respectively), indicating a mixed-order transition having the same universality class as percolation on abstract interdependent networks (8).

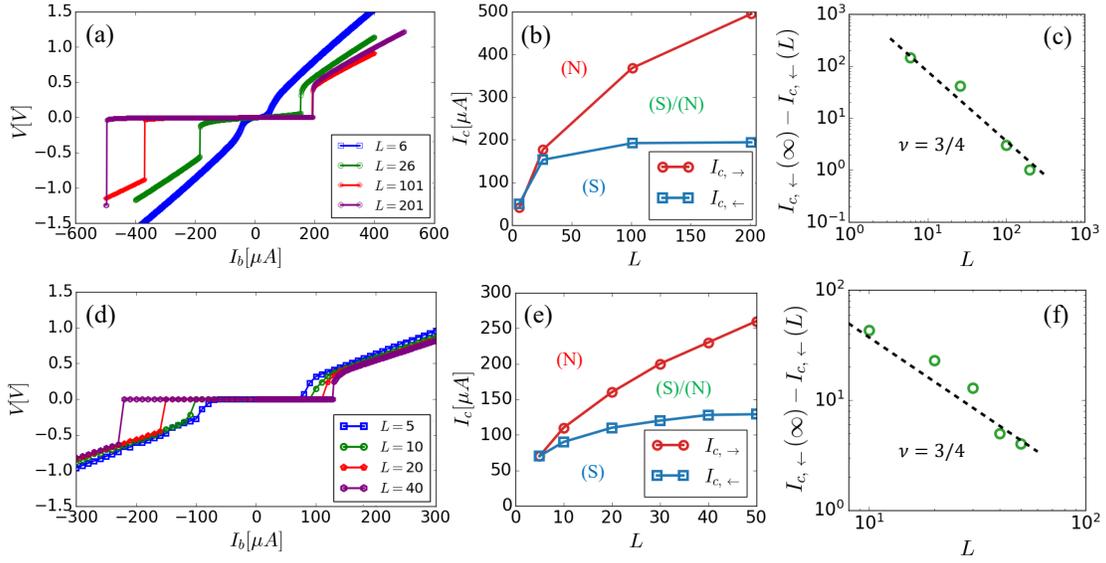

**Figure 2**: **System size dependence.** (a) Experimental results (top networks) for the $I-V$ curves of interdependent superconducting networks for different system sizes. Small systems experience continuous phase transition like a single layer (*7*) while for large systems abrupt transition is observed. The asymmetry in the characteristics of the $I-V$ curves is the result of the different critical points for increasing and decreasing currents, i.e., the hysteresis. (b) This behavior is reflected by the critical currents. For small systems, identical critical points are shown for both increasing and decreasing current $I_{c,\rightarrow} = I_{c,\leftarrow}$. Larger systems experience an abrupt transition and hysteresis appears with $I_{c,\rightarrow} > I_{c,\leftarrow}$. (c) The correlation length exponent $\nu = 3/4$ is evaluated using the scaling of the critical point with the system size (Eq. (2)) as predicted for interdependent systems (*26*). (d)-(f) The theory based on numerical results for the iterative solution of the coupled Kirchhoff equations supports the experimental findings shown in (a)-(c) respectively.

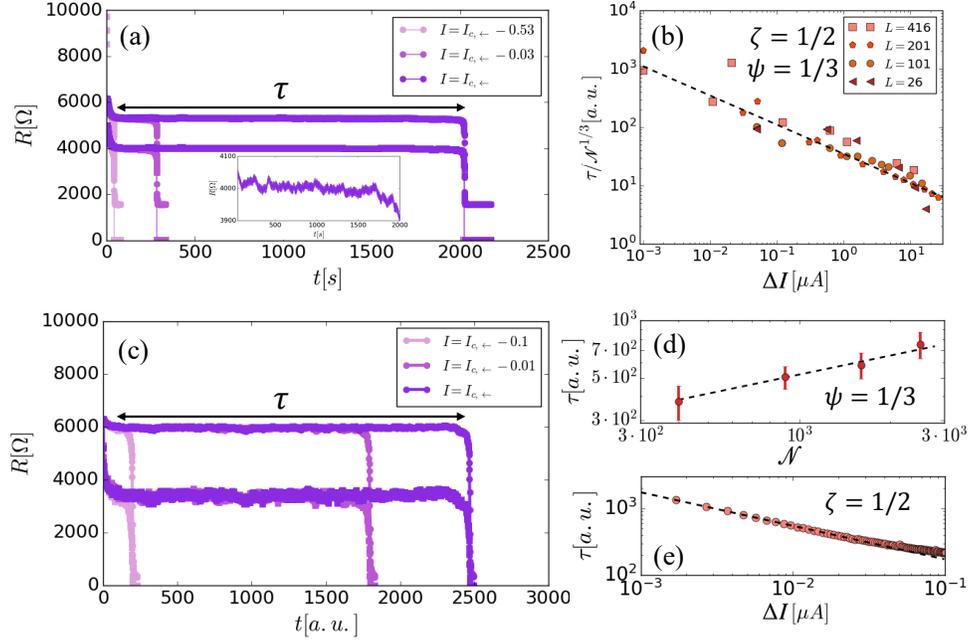

**Figure 3**: **Plateau.** (a) The resistance of an $L = 416$ ISN is measured experimentally as a function of time (in seconds) during the abrupt transition from the mutual N-state to the mutual S-state for different bias currents $I_b \leq I_c$, and a long-term plateau is observed. The time duration of the plateau $\tau$ decreases as the current departs from $I_c$. The inset zooms in on the plateau regime of the top network showing that the resistance monotonically decreases but is affected by thermal fluctuations. (b) The plateau timescale $\tau$ follows the scaling with the distance from criticality in Eq. (3) with $\zeta = 1/2$ and the scaling with the system size in Eq. (4) with $\psi = 1/3$. Both exponents are similar to those found for percolation on abstract interdependent networks (*27*). (c) Long-lived plateau is observed theoretically by solving numerically the Kirchhoff equations of the thermally interdependent superconducting networks. (d) In the numerical simulations one can see that the plateau time scale at the critical current $I_c$ increases with the system size according to Eq. (4) with $\psi = 1/3$ and (e) decreases with the distance from criticality according to Eq. (3) with $\zeta = 1/2$, in excellent agreement with experiment shown in (b).

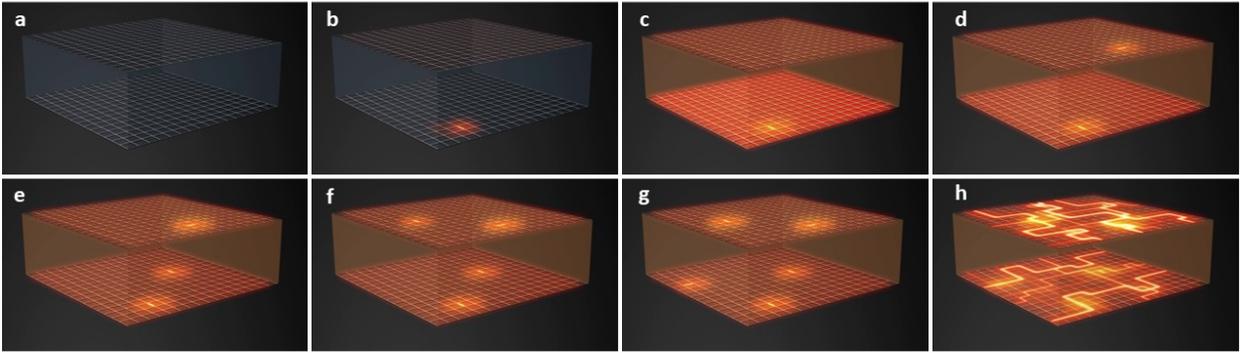

**Figure 4**: **Illustration of the dynamics of the cascading mechanism for a heating process during the abrupt transition.** (a) An ISN system is stabilized at $T < T_c$ and $I < I_c$ where all segments are superconducting and there is no thermal dissipation in the system. (b) At time $t = 0$ the current is switched to $I_b > I_c$ so that one segment switched to the N state and (c) dissipates heat to the entire networks system. (d) This causes a random segment in the second network to switch to the N phase and to dissipate heat. (e-g) This feedback process of one segment creating a phase change in a random segment in the other network, continues spontaneously back and forth for a macroscopic time until (h) an abrupt transition occurs and both networks become normal.

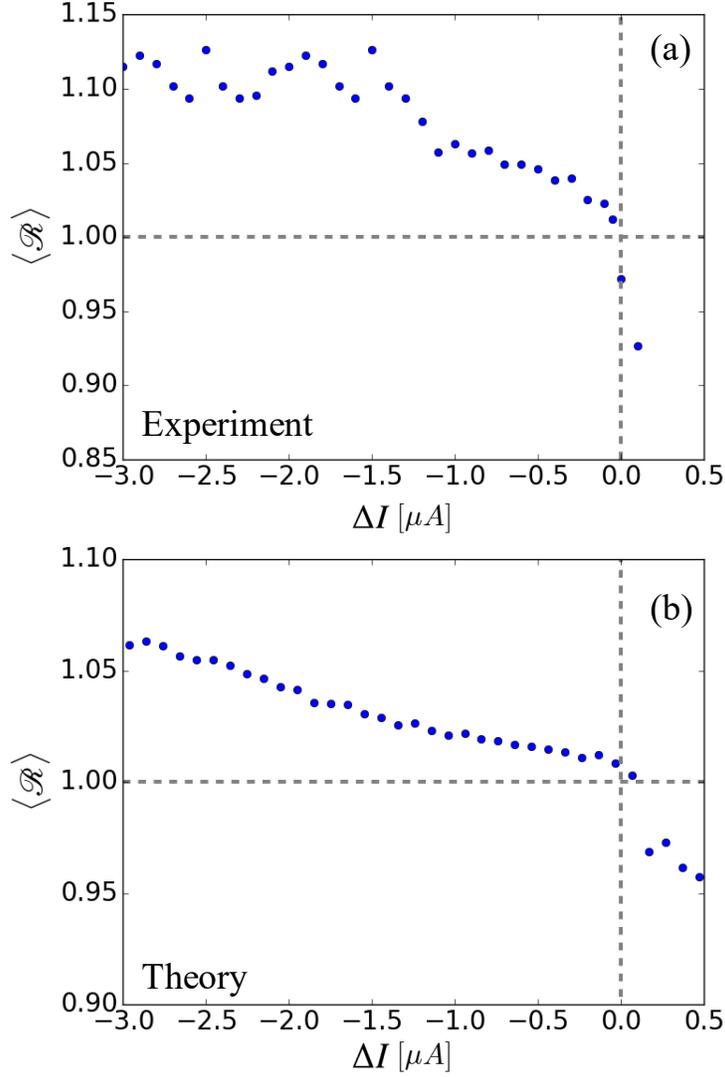

**Figure 5**: **Branching factor** (a) Experimental (top network of an $L = 416$ ISN) and (b) theoretical measurements of the average branching factor during the cascading near criticality of the abrupt transition. These are extracted from resistance versus time curves shown in Fig. S3. Above the critical point ($\Delta I > 0$), the average branching factor $\langle \mathcal{R} \rangle$ is smaller than one while below it ($\Delta I < 0$), the average branching factor is larger than one as hypothesized. Note also that $\langle \mathcal{R} \rangle$ approaches 1 as $I$ approaches $I_c$ from both directions. Exactly at the critical point ($\Delta I = 0$) a critical branching factor of $\langle \mathcal{R} \rangle_c = 1$ is observed, as hypothesized and explained in the text.

# Acknowledgments

We thank N. Shnerb for helpful discussions.

**Funding:** S.H. acknowledges the support of the Israel Science Foundation (Grant No. 189/19), the Binational Israel-China Science Foundation Grant No. 3132/19, the EU H2020 project RISE (Project No. 821115), the EU H2020 DIT4TRAM, EU H2020 project OMINO (Grant No. 101086321) for financial support. B.G. acknowledges the support of the Fulbright Postdoctoral Fellowship Program. I.V. Y.S. and A.F. acknowledge support from the Israel Science Foundation (ISF) Grants No. 3053/23 and No. 1499/21.

**Authors contribution:** I.V. and Y.S. prepared the samples and performed the experiments. B.G., I.B., and N.Y. performed the theoretical calculations and data analysis. AF and SH initiated and supervised the project. B.G., A.F., and S.H. designed the research and wrote the manuscript.

**Competing interests:** There are no competing interests to declare.

**Data and materials availability:** Source codes can be freely accessed at the GitHub repository: https://github.com/BnayaGross/Microscopic-mechanism-of-interdependent-SC-networks. Experimental measurements data is attached to Supplementary Materials.

# Supplementary materials

Materials and Methods

Supplementary Text

Figs. S1 to S4

Data S1 to S4

Video S1